\documentclass[aps,prb,twocolumn]{revtex4}
\usepackage{amsmath, amsthm, amssymb,graphicx}
\begin{document}
\title{Spontaneous Symmetry Breaking and Decoherence in Superconductors}
\author{Jasper van Wezel and Jeroen van den Brink}

\affiliation{
Institute-Lorentz for Theoretical Physics,  Universiteit  Leiden,
P.O. Box 9506, 2300 RA Leiden, The Netherlands}
\date{\today}

\begin{abstract}
We show that superconductors have a thin spectrum associated with spontaneous symmetry breaking similar to that of antiferromagnets, while still being in full agreement with Elitzur's theorem, which forbids the spontaneous breaking of local (gauge) symmetries. This thin spectrum in the superconductors consists of in-gap states that are associated with the spontaneous breaking of a \emph{global} phase symmetry. In qubits based on mesoscopic superconducting devices, the presence of the thin spectrum implies a maximum coherence time which is proportional to the number of Cooper pairs in the device. Here we present the detailed calculations leading up
to these results and discuss the relation between spontaneous symmetry breaking in superconductors and the Meissner effect, the Anderson-Higgs mechanism and the Josephson effect. Whereas for the Meissner effect a symmetry breaking of the phase of the superconductor is not required, it is essential for the Josephson effect.
\end{abstract}

\maketitle

\section{Introduction}
Recently we have shown that spontaneous symmetry breaking imposes a
fundamental limit to the time that a large spin system can stay quantum
coherent. This coherence timescale is $t_{spon} \simeq 2 \pi N \hbar / \left(
k_B T \right)$, given in terms of the number of microscopic degrees of freedom
$N$, temperature $T$ and the constants of Planck ($\hbar$) and Boltzmann
($k_B$).\cite{vanWezel05, vanWezel06,vanWezel07,VanWezel06b,vanWezelSC,Rycerz06,Yuan07}
The timescale $t_{spon}$ is expected to be a
universal timescale associated with all forms of spontaneous symmetry
breaking, since it does not depend on any of the model parameters that were
needed to derive it. In this paper we will show in detail how superconducting systems spontaneously break their phase
symmetry and that they have a thin spectrum associated with this.\cite{vanWezelSC} The thin spectrum subtly influences the dynamics of the superconductor when it is used as a qubit. The resulting maximum coherence time is again given by the universal expression $t_{spon}$.

It has been proven already three decades ago by Elitzur that local (gauge) symmetries cannot be broken spontaneously without invoking an explicitly asymmetric gauge fix.\cite{Elitzur75} Also, it has been argued recently that because the local gauge symmetry in superconductors cannot be broken spontaneously, the order should be of a purely topological nature, and that the low energy properties of the superconducting state are determined solely by its topological structure.\cite{Hansson04} At first sight then, the claim that a superconductor possesses states related to spontaneous symmetry breaking that are at very low energy and within the superconducting energy gap, might come as a surprise.

However, it is well known that the superconducting ground state is characterized by a definite phase and a corresponding uncertainty in the number of Cooper pairs.
%
%
For a piece of superconducting material the realization of one specific phase is but one choice out of a manifold of equivalent possibilities. Its phase having just one specific value therefore has to come about by spontaneous symmetry breaking. This symmetry breaking in the thermodynamic limit requires the existence of a so called thin spectrum of total phase states whose energies all collapse onto the groundstate energy in the thermodynamic limit. Such a spontaneous symmetry breaking is {\it not} at variance with Elitzur's theorem because the symmetry that is broken in a superconductor is a \emph{global} U(1) phase symmetry. The resulting superconducting state is still manifestly invariant under local gauge transformations.

To clearly illustrate these points we will first discuss the superconducting state of an array of Josephson junctions. In this array the non-commutativity of number and phase variables straightforwardly gives rise  to spontaneous symmetry breaking and to a thin spectrum. Symmetry breaking in this superconductor turns out to be exactly analogous to the case of quantum crystals or magnetic systems.\cite{vanWezel06} After that
we will switch to a microscopic strong coupling model of superconductivity in which the
role of gauge symmetry can be more clearly discussed. We will then use this
model to describe a Cooper-pair box qubit and show that the presence of the
thin spectrum leads to a maximum coherence time $t_{spon}$ of the
qubit, which is of the order of milliseconds. Finally we will show how the description of the thin spectrum can also be
incorporated into the familiar BCS description of superconductivity, and
comment on the application to different types of qubits.

\section{Josephson Junction Array}

It is well known that an array of superconducting islands coupled together by
Josephson junctions can undergo a (quantum) phase transition from an
insulating state to a superconducting state.\cite{Simanek79, Efetov80,
  Bradley84, Fisher88, Wallin94, Sondhi97} The description of a
superconductor as an array of Josephson junctions is particularly useful to us
here because it naturally focuses the attention on the the conjugate
variables number and phase.  The Hamiltonian for the Josephson junction array is given by
\begin{eqnarray}
H^{JJ} &=& \sum_{j,\delta} \left[ \frac{E_C}{2} n_j^2 - E_J \cos\left(\theta_j -
  \theta_{j+\delta}  \right) \right].
\label{H:JJ}
\end{eqnarray}
Here $\theta_j$ represents the phase of the superconducting orderparameter of superconducting island $j$, while $n_j$ gives the number of Cooper pairs above average, and $\delta$ connects neighboring sites. The
charge or number operator $n_j = -i (\partial/\partial \theta_j)$ is the variable conjugate to the
phase, and can be written in terms of the voltage $V$ and the capacitance $C$
of the Josephson junctions as $n_j =(C/2e) V_j$. The coupling constants are
the charging energy $E_C$ and the Josephson coupling energy $E_J$, which are chosen to lie well within the superconducting regime.

The phase $\theta$ in this description can be thought of as the phase of the
Ginzburg-Landau wavefunction for the superconducting island, or equivalently
as the phase describing the perfectly ordered BCS state $\left| \theta \right>
= \prod_k \left(\left|u_k \right| + \left|v_k \right| e^{i \theta}
c^{\dagger}_k c^{\dagger}_{-k} \right) \left| \text{vac}
\right>$.\cite{Tinkham} This phase is not measurable as such, but a
difference in phase across a Josephson junction causes a supercurrent $J=J_C
\sin \left( \theta_j - \theta_{j+\delta} \right)$, and therefore phase differences are measurable. The condition of measurability implies the gauge independence of these quantities, because a gauge transformation by definition cannot alter the outcome of any experiment. The total phase is both unmeasurable and a gauge dependent quantity.

The thin spectrum of the Josephson junction array consists of the infinite wavelength
part of the Hamiltonian~Eq.\eqref{H:JJ}, because exactly at $k=0$ the Bogoliubov
transformation that would diagonalize the Hamiltonian turns out to be
singular.\cite{vanWezel06} This zero wavenumber
part of $H^{JJ}$ which describes the collective behavior of the system as a whole, is given by
\begin{eqnarray}
H_{{\bf k}=0}^{JJ} = \frac{E_C}{2 N} n_{tot}^2,
\label{H0}
\end{eqnarray}
where $N$ is the total number of superconducting islands, and $n_{tot} \equiv
\sum_j n_j$ is the charge of the total network of Josephson junctions. To see
how the array can spontaneously break its total phase symmetry we should add a
symmetry breaking field to the collective Hamiltonian. We cannot simply add a
term which involves the bare total phase $\theta_{tot}$, because that total
phase is not a gauge independent, measurable quantity. Instead we can look at
the difference of phase between the Josephson junction array and some given
reference superconductor. In the end we will let the strength of the
symmetry breaking field go to zero, or equivalently move the reference
superconductor away to infinity. The Hamiltonian including the symmetry
breaking field thus becomes
\begin{eqnarray}
H^{JJ}_{SB} = \frac{E_C}{2 N} n_{tot}^2 - B \cos\left( \theta_{tot} -
\theta_{ref} \right).
\label{HjjSB}
\end{eqnarray}
For small values of $\Delta \theta_{tot} \equiv \theta_{tot} - \theta_{ref}$ we can expand the cosine to
quadratic order and then the Hamiltonian reduces to a harmonic oscillator with
well known solutions in terms of Hermite polynomials, in exact analogy to the case of spontaneous symmetry breaking in quantum crystals and antiferromagnets. Using these Hermite polynomials, it is easy
to show that indeed the Josephson junction array can spontaneously break the
rotational symmetry of its total phase by looking at its fluctuations in the
limit of disappearing symmetry breaking field and infinite number of
superconducting islands:
\begin{eqnarray}
f^2 \equiv \left< \left(\Delta \theta_{tot}\right)^2 \right> - \left<
\Delta \theta_{tot} \right>^2 &\propto& \frac{1}{\sqrt{N B}} \nonumber \\
\lim_{N \to \infty} \lim_{B \to 0} f^2 &\to& \infty
\nonumber \\
\lim_{B \to 0} \lim_{N \to \infty} f^2 &\to& 0.
\label{limits}
\end{eqnarray}
Clearly the fluctuations in the total phase disappear in the thermodynamic
limit even if only an infinitesimal symmetry breaking field is present.

The symmetry broken state that is formed in that limit has a well defined total phase,
and must thus be in a superposition of many different total number
states. These total number states were precisely the eigenstates of the
collective Hamiltonian~Eq.\eqref{H0}, which we identified as being the thin
spectrum of the Josephson junction array. The symmetry broken Hamiltonian also
has a tower of low lying states that form a sort of dual thin spectrum which
consists of all the total phase states necessary to build a state with a
fixed total number of Cooper pairs. Notice that the thin spectrum states must be observable states, because the description of the collective dynamics in Hamiltonian~Eq.\eqref{HjjSB} is still manifestly gauge invariant. This also implies that the symmetry breaking which we have just described is \emph{not} the breaking of a local gauge symmetry. Only the U(1) symmetry of the global total phase is spontaneously broken, and even then only in the sense that its fluctuations disappear in the thermodynamic limit, so that its value relative to that of some other, external superconductor will be fixed.

The fact that we needed to introduce an external superconductor as a {\it deus ex machina} to fix the phase of our Josephson junction array should come as no surprise. The situation is in fact precisely analogous to that of breaking the translational symmetry of a crystal. In that case one can only assign a definite value to the position of the symmetry-broken crystal by measuring the distance of its center of mass to some external reference point (which in that case can the observer himself). The position of the entire system of crystal and observer together is still completely arbitrary (or at least unmeasurable for the observer), even in the symmetry-broken state.

\section{Local Pairing Superconductor}
From the previous section it is clear that that the non-commutativity of number and phase naturally gives rise to the presence of a thin spectrum in a superconducting system. Now we will examine how the superconducting orderparameter comes about by spontaneous symmetry breaking in the first place and how this is related to gauge symmetry. This relation was not visible in the context of a Josephson junction array, because there we started out with islands that were postulated to be in a superconducting state. That way we could describe the whole system with an effective Hamiltonian that only consisted of observables related to the macroscopic properties of the superconducting state.

For a more general description of superconductivity we start out with a microscopic Hamiltonian for a single superconductor that incorporates the effects of the gauge field. The simplest such model is the extensively studied local pairing, negative-$U$ Hubbard model\cite{Micnas90,Zaanen96}
\begin{eqnarray}
H &=& \frac{1}{2} \sum_{j,\delta,\sigma} \left( t_j^{\delta}
c^{\dagger}_{j+\delta,\sigma} c^{\phantom \dagger}_{j,\sigma} + \left(
t_j^{\delta} \right)^* c^{\dagger}_{j,\sigma} c^{\phantom
  \dagger}_{j+\delta,\sigma} \right) \nonumber \\
&&\hspace{12pt} -\left| U \right| \sum_j n_{j,\uparrow} n_{j,\downarrow}.
\label{Hzen}
\end{eqnarray}
Here $c^{\dagger}_j$ creates an electron on site $j$, $\delta$ connects
neighboring sites and $n_j$ counts the number of electrons. The reason to consider this local pairing model rather than for example the BCS model for superconductivity is the fact that this model is explicitly gauge invariant, while the BCS model is not. From the symmetry point of view, the models are the same: there is no phase transition in going from weak to strong coupling superconductivity, only a cross-over.
If we parametrize the hopping in terms of a
uniform amplitude and a bond dependent phase as $t^{\delta}_j=t e^{i
  \psi^{\delta}_j}$, then minimal coupling allows us to identify the phase
of the hopping parameter with the electromagnetic vector potential integrated along the bond under
consideration, so that $\psi^{\delta}_j=\frac{e}{\hbar c} \int_j^{j+\delta}
A^{\delta}(t) dt$. Thus the Hamiltonian is invariant under the gauge transformation
\begin{eqnarray}
c^{\dagger}_j &\to& e^{i \frac{e}{\hbar c} f(j)} c^{\dagger}_j, \nonumber \\
{{\bf A}}(j) &\to& {{\bf A}}(j) + {{\bf \nabla}} f(j),
\end{eqnarray}
which immediately implies
\begin{eqnarray}
\psi^{\delta}_j &\to&
\psi^{\delta}_j +\frac{e}{\hbar c} \left[f(j+\delta) - f(j)\right].
\end{eqnarray}

We focus on the strong coupling limit where $U \gg t$, so that we only need to consider the physics of the lower Hubbard sector. On each site there will thus be either a pair of electrons or no electrons at all. Single electron excitations are only virtually allowed and give rise to pair-pair interactions. The effective low energy Hamiltonian is given by a second order perturbation expansion in the hopping and can be written in terms of pseudospin operators that are defined by
\begin{eqnarray}
S_j^{+} &=& c^{\dagger}_{j,\uparrow} c^{\dagger}_{j,\downarrow} \nonumber \\
S_j^z &=& \frac{1}{2}\left(n_{j,\uparrow} + n_{j,\downarrow}-1\right).
\end{eqnarray}
The $z$ projection of the pseudospin measures the local electron density, while the $xy$ components provide the dynamics of the Cooper pairs. Adding a chemical potential $\mu$ that determines the overall electron density and thus explicitly breaks the electron-hole symmetry, we find the effective Hamiltonian
\begin{eqnarray}
H_{eff} &=& \frac{J}{2} \sum_{j,\delta} \left[ e^{i 2 \psi_j^{\delta}}
  S_j^+ S_{j+\delta}^- +e^{-i 2 \psi_j^{\delta}}
  S_j^- S_{j+\delta}^+ \right] \nonumber \\
&+& J \sum_{j,\delta} \left[
  S_j^z S_{j+\delta}^z -\frac{1}{4} \right] - h \sum_{j} \left[ S_j^z +\frac{1}{2} \right].
\label{Heff}
\end{eqnarray}
Here $J$ is defined to be $2 t^2/|U|$, and $h \equiv |U|-2\mu$
determines the overall electron density. Away from half filling the global
SU(2) symmetry of the Hamiltonian is manifestly broken, and what remains is the U(1)
symmetry that describes rotations around the $z$-axis. It is the spontaneous breaking of this global U(1) symmetry that will yield the superconducting state.

\section{Meissner Effect and Anderson-Higgs Mechanism}
Before we discuss the actual spontaneous symmetry breaking and the thin spectrum associated with it, we will show that already on the semi-classical level, the model~Eq.\eqref{Heff} can be seen to expel magnetic field lines from its ground state and to give propagating electromagnetic modes in its bulk a finite effective mass (the Meissner effect and Anderson-Higgs mechanism).

To find a semiclassical description for the groundstate of the $S=1/2$ pseudospin Hamiltonian $H_{eff}$, we introduce generalized coherent states of the form
\begin{eqnarray}
\left| \Psi_{class} \right> &=& \prod_j \left( e^{-i \frac{\phi_j}{2}} \sin \left(
\frac{\theta_j}{2} \right) \right. \nonumber \\
&& \left. \hspace{18pt} + e^{i \frac{\phi_j}{2}} \cos \left( \frac{\theta_j}{2} \right)
c^{\dagger}_{j,\uparrow} c^{\dagger}_{j,\downarrow} \right) \left|vac\right>.
\label{psi}
\end{eqnarray}
In this expression the angles $\phi_j$ and $\theta_j$ are the Euler angles
which describe the classical vectors that replace the quantum spins in the
classical state. To find the semiclassical groundstate energy we need to
minimize the expectation value of $H_{eff}$ in the generalized coherent state with respect to the orientations of the classical
spin-vectors. It is easy to check that the classical energy can be minimized by first fixing the azimuthal angles $\theta_j$ uniformly throughout the system, after which the energy up to a constant is given by
\begin{eqnarray}
E_{class}
 &\simeq& J \rho \sum_{j,\delta} \cos \left( 2 \psi_j^{\delta} + \phi_j - \phi_{j+\delta} \right),
\label{Eclass}
\end{eqnarray}
where $\rho$ is a constant set by the optimized value of $\theta$.
In this expression the global SU(2) rotational symmetry has already been broken explicitly by the effect of the field $h$ (which fixed the azimuthal angles). What is left is the in-plane U(1) rotational symmetry of the polar angles $\phi_j$.
The classical state with lowest energy links the values of these polar angles to the bond variables $\psi_j^{\delta}$. These variables are in turn connected to the electromagnetic vector potential. One finds that the condition for the angles $\phi_j$ that minimizes the energy is
\begin{eqnarray}
\bar{A}^{\delta}_j = \frac{\hbar c}{2 e} \frac{\phi_{j+\delta} - \phi_j +
  \pi}{a},
\label{MeissnerDisc}
\end{eqnarray}
where $\bar{A}^{\delta}_j$ is the average value of the vector potential along the
bond. At distances much larger than the lattice spacing $a$ this expression
becomes
\begin{eqnarray}
{\bf A}\left({\bf r}\right) = \frac{\hbar c}{2 e} {\bf \nabla}
\phi\left({\bf r}\right).
\label{MeissnerCont}
\end{eqnarray}

The classical groundstate is thus a state in which the electromagnetic potential is proportional to the gradient of the scalar field $\phi$, which of course immediately implies that the rotation of $\bf A$ will vanish, and thus that the condensate does not allow any magnetic field to penetrate its bulk --a clear indication that the semiclassical groundstate is indeed a superconducting state.

For a full description of the Meissner effect, however, it is not enough to show that the semiclassical ground state does not contain a magnetic field. One also needs to demonstrate that the superconducting classical state will also actively expel externally applied electromagnetic fields; i.e. that the electromagnetic excitations in the system are gapped and massive. To do so we add an external electromagnetic field to the semiclassical energy density:
\begin{eqnarray}
E_{class}
 &=& \frac{J \rho}{N} \sum_{j,\delta} \cos \left( \frac{2 e a}{\hbar c} \bar{A}^{\delta}_j + \phi_j - \phi_{j+\delta} \right) \nonumber \\
&& +\frac{1}{2} \left({\bf \nabla} \times {\bf A}\right)^2 +\frac{1}{2}\left(\dot{{\bf A}}\right)^2.
\label{EclassEM}
\end{eqnarray}
This expression can be simplified by introducing a new vector field ${\bf A}' \equiv {\bf A} - {\bf \nabla} \phi$. Because the newly defined field ${\bf A}'$ is formally equivalent to a gauge-deformed version of the electromagnetic field ${\bf A}$, we can be sure that the electromagnetic energy $E^2+B^2$ looks the same in terms of ${\bf A}'$ as it did in terms of ${\bf A}$. If we take the continuum limit and expand the cosine to second order, we thus find
\begin{eqnarray}
E_{class}
&\simeq& J' \left( {\bf A}' \right)^2 +\frac{1}{2} \left({\bf \nabla} \times {\bf A}'\right)^2 +\frac{1}{2}\left(\dot{{\bf A}}'\right)^2
\label{Higgs}
\end{eqnarray}
Note that both expressions for the energy given above are fully gauge invariant. Using the Hamilton equations it can immediately be checked that the latter expression for the classical energy yields only massive propagating modes in terms of the field ${\bf A}'$. Due to this Anderson-Higgs mechanism, the physical excitations of the system --which are combined modes of the electromagnetic field and phase degree of freedom-- have a finite energy gap. This prevents an external electromagnetic field to penetrate the bulk of the superconductor.

\section{Thin Spectrum}
We have seen in the previous section that the semicalssical groundstate of the tight binding negative-$U$ Hubbard model  displays the Meissner effect and generates a mass for the electromagnetic modes via the Anderson-Higgs mechanism. In fact in that semiclassical description the question as to whether the global U(1) symmetry was broken or not never arose. We thus conclude that the Meissner effect and Anderson-Higgs mechanism occurs regardless of whether the superconductor has a well-defined total phase.\cite{Greiter,Penrose,Yang} The situation is similar to that in antiferromagnets, where long range antiferromagnetic correlations exist both in the symmetric singlet ground state and in the symmetry broken N\'eel state.

\begin{figure}
\includegraphics[width=.8\columnwidth]{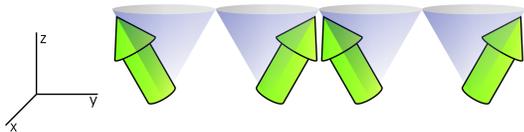}
\caption{A schematic one-dimensional representation of the classically realized state of the tight binding superconductor. The arrows are a classical cartoon for the transformed pseudospins ${\bf \sigma}$.}
\label{canted}
\end{figure}

To see the effects of spontaneous symmetry breaking in the negative-$U$ Hubbard model we need to describe the formation of the symmetry broken ground state in a more analytical manner by studying the exact eigenstates of the collective part of the Hamiltonian without resorting to semi-classics, just as we did for the Josephson junction array. The difficulty in such a global description will be to correctly account for the gauge field, which can fluctuate locally. To circumvent this
problem we introduce transformed pseudospins, analogous to what is done in the weak coupling theory\cite{Anderson58,Micnas90}
\begin{eqnarray}
\sigma_j^+ &=& e^{-2i \sum_{j'=0}^j \psi_{j'}^{\delta}} S^+_j \nonumber \\
\sigma_j^z &=& S_j^z.
\label{trafo}
\end{eqnarray}
The summation in the exponent is over some path connecting position $j$ to
some origin $j=0$. For simplicity we will assume the applied external
magnetic field to be zero from here on.
Notice that the individual transformed pseudospin operators of equation~Eq.\eqref{trafo} are not gauge invariant.
Their purpose is to transform the local gauge
transformations of the actual pseudospins ${\bf S}$ into a
global transformation of the new pseudospins ${\bf \sigma}$:
\begin{eqnarray}
\sigma^+_j &\to& e^{-2i \sum_{j'=0}^j \left[ \psi_{j'}^{\delta} + \frac{e}{\hbar c}
  \left( f(j'+\delta) -f(j')\right)\right]} e^{2i \frac{e}{\hbar c} f(j)} S^+_j \nonumber \\
&=& e^{2i \frac{e}{\hbar c} f(0)} \sigma^+_j \equiv e^{i \psi_0} \sigma^+_j \nonumber \\
\sigma^z_j &\to& \sigma^z_j.
\end{eqnarray}
A gauge transformation therefore corresponds to a rotation along the $z$-axis of all pseudospins by the same angle. It is this global character of the gauge transformations on the transformed pseudospins that allows us to switch to a description of just the collective behavior of the system without invoking any specific gauge choice. In terms of the transformed pseudospins the effective low energy Hamiltonian~Eq.\eqref{Heff} becomes
\begin{eqnarray}
H_{eff} = J \sum_{j,\delta} {\bf \sigma}_j^{\phantom z} \cdot {\bf \sigma}_{j+\delta}^{\phantom z} - h
\sum_j \sigma_j^z.
\end{eqnarray}
The parameter $J$ is positive because the electron pairs repel each other. This Hamiltonian therefore describes an antiferromagnetic interaction between neighboring pseudospins, and an overall, uniform magnetic field. The classical state that we expect to find in terms of the pseudospins ${\bf \sigma}$ is therefore a canted
antiferromagnet. That is, an antiferromagnet in which all spins are uniformly
canted out of the $z=0$ plane, but in which the $xy$-projections still form an
antiferromagnetic pattern (see figure~\ref{canted}). As we noticed before, the canting of the spins which
breaks the full SU(2) symmetry down to U(1) is done explicitly by the field $h$, while the
breaking of the in-plane U(1) symmetry into an antiferromagnetic structure
will have to be realized through spontaneous symmetry breaking.

The thin spectrum of the Hamiltonian $H_{eff}$ consists of the states necessary to construct the symmetry broken classical state. These thin spectrum states describe the dynamics of the superconductor as a whole, and as before they can be found at the singular points of the Bogoliubov transformation that diagonalizes the quadratic part of the Hamiltonian.\cite{vanWezel06} In the antiferromagnet both the point $k=0$ and the point $k=\pi$ are singular. The resulting collective part of the Hamiltonian is given by
\begin{eqnarray}
H_{coll} = \frac{4 J}{N} {\bf \sigma}_A^{\phantom z} \cdot {\bf \sigma}_B^{\phantom z} - h \sigma_{tot}^z,
\label{coll}
\end{eqnarray}
where ${\bf \sigma}_{A,B}$ denotes all spins on the $A,B$ sublattice and ${\bf \sigma}_{tot}$ is the combination of all spins on the entire lattice. The modes that form the thin spectrum are not affected by the Anderson-Higgs mechanism that we discussed before because they are zero-wavelength, global excitations which only affect the system as a whole. The coupling of the phase degrees of freedom and the electromagnetic field in the Anderson-Higgs description exists only at finite wavelength, as can be easily seen in equation~Eq.\eqref{EclassEM} because a global transformation $\phi_i \rightarrow \phi_i + \delta \phi$ for all $i$ leaves the coupling of the phase to the electromagnetic field invariant. The Hamiltonian $H_{coll}$ therefore captures all the relevant collective, low energy behavior of the model.

The collective Hamiltonian~Eq.\eqref{coll} is just a Lieb Mattis model in a uniform magnetic field,\cite{vanWezel06,Lieb62,Kaplan90,Kaiser89} and the eigenstates are readily identified as the states labeled by the quantum numbers $\sigma_A^{\phantom z}, \sigma_B^{\phantom z}, \sigma_{tot}^{\phantom z}$ and $\sigma_{tot}^z$. The difference between this collective model and the one describing the spontaneous symmetry breaking in antiferromagnets is the field $h$, which reduces the symmetry from SU(2) to U(1). The ground state has maximum total spin on both the $A$ and $B$ sublattice, and has a total spin that is equal to its $z$-projection of  $\sigma_{tot}^z=\sigma_{tot}^{\phantom z}=(h N)/(4 J)$. Excitations of the quantum numbers $\sigma_A^{\phantom z}$ and $\sigma_B^{\phantom z}$ are gapped with an energy $J$ from the groundstate, because of the infinitely long range of the interactions in $H_{coll}$. We will henceforth set these quantum numbers to their maximum value and only consider the low energy excitations which describe the behavior of the entire system as a whole. We can relabel the eigenstates by introducing
\begin{eqnarray}
\sigma_{tot}^{\phantom z} &=& \bar{\sigma} + n \nonumber \\
\sigma_{tot}^{z} &=& \bar{\sigma} + n - y.
\end{eqnarray}
Here $\bar{\sigma}$ is the groundstate value for the $z$-projection of the pseudospin: $\bar{\sigma}=(h N)/(4 J)$. In terms of the quantum numbers $n$ and $y$, the effective Schr\"{o}dinger's equation for the excitations of the Lieb-Mattis Hamiltonian $H_{coll}$ becomes
\begin{eqnarray}
H_{coll} \left| n,y \right> = \left( E_{coll}^0 + h y + \frac{2 J}{N} n^2 \right) \left| n,y \right>.
\end{eqnarray}
From this equation it is clear that the excitations labeled by $n$ will play
the role of the thin spectrum for the tight binding superconductor. It can be
easily checked that indeed the contribution of these states to the partition
function vanishes in the thermodynamic limit. The excitation labeled by $y$ on
the other hand is a collective excitation that changes the $z$-projections of
all pseudospins and costs an energy proportional to the chemical potential to
excite. This is, in other words, the quantum number that determines the
average total number of Cooper pairs in the superconductor.

\begin{figure}
\includegraphics[width=.5\columnwidth]{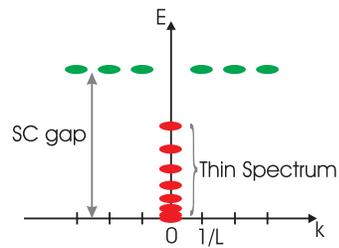}
\caption{Schematic representation of the dispersion relation of the low-energy, low-momentum states of a finite size superconductor: the states at finite $k$ are gapped, while the thin spectrum states at $k=0$ lie within the gap.}
\label{spectrumplot}
\end{figure}

\section{Breaking the Symmetry}
To study the spontaneous symmetry breaking of $H_{coll}$ we will have to introduce a symmetry breaking field, which we will send to zero again at the end of the calculation. In analogy to the symmetry breaking term that we used in the Josephson junction array, we will again introduce a second external superconductor which is weakly coupled to the first, and let the coupling tend to zero:
\begin{eqnarray}
H_{coll}^{SB}=H_{coll} + T \left(\Sigma_A^+ \sigma_A^- + \Sigma_B^+ \sigma_B^- + \text{H.c.} \right) + H_{ext}
\label{eq:Hsb}
\end{eqnarray}
Here ${\bf \Sigma}_{A,B}$ denotes the pseudospin operators in the external superconductor and $H_{ext}$ describes its dynamics. For simplicity we assume that both the Hamiltonian $H_{coll}$ and $H_{ext}$ are given by the collective model of equation~Eq.\eqref{coll}. Notice that the coupling Hamiltonian is still completely invariant under local gauge transformations which act on the pseudospins ${\bf \sigma}_{A,B}$ as well as on ${\bf \Sigma}_{A,B}$. The symmetry that is broken in equation~Eq.\eqref{eq:Hsb} is the U(1) phase symmetry that rotates all pseudospins ${\bf \sigma}$, but keeps the pseudospins ${\bf \Sigma}$ fixed: the phase {\em difference} between the superconductors can acquire a finite expectation value in the symmetry broken state, without violating Elitzur's theorem.

To explicitly see the influence of the external superconductor, we would liketo use the total phase of the system $H_{ext}$ as a reference point for measuring the phase of $H_{coll}$. To do so, we need to temporarily abandon the manifest gauge invariance of our description, and assume that the total-phase symmetry of $H_{ext}$ has already been broken. This allows $H_{ext}$ to generate a symmetry breaking field for the $\sigma$-pseudospins subsystem. Using for this (mean) field the expectation values $\left< \Sigma_A^x \right> = - \left< \Sigma_B^x \right> = N_{\Sigma}/2$ and $\left< \Sigma_A^y \right> = \left< \Sigma_B^y \right> = 0$, the effective Hamiltonian for the original superconductor reduces to
\begin{eqnarray}
H_{coll}^{SB} = \frac{4 J}{N} {\bf \sigma}_A^{\phantom z} \cdot {\bf \sigma}_B^{\phantom z} - h \sigma_{tot}^z - B \left(\sigma_A^x -\sigma_B^x \right),
\label{HcollSB}
\end{eqnarray}
with $B=T N_{\Sigma}$, and $N_{\Sigma}$ the number of pseudospins in the external superconductor. Notice that the symmetry breaking term in this Hamiltonian is \emph{not} gauge invariant. As a gauge transformation corresponds to a uniform rotation of all spins on the entire lattice around the $z$ axis, any explicit choice for the direction of $B$ along a particular axis in the $xy$-plane is connected  by a gauge transformation to any other directions in the plane. By fixing the total phase of the external superconductor to lie along the $x$-axis in equation~Eq.\eqref{HcollSB}, we have thus implemented a specific gauge choice, and we will have to check afterward if the conclusions based on calculations in this particular gauge fix are robust under gauge transformations.

The matrix elements of the symmetry breaking field in the basis $\left| n,y \right>$ can be computed by performing a sum over Clebsh-Gordon coefficients.\cite{vanWezel06} In the limit $(n,y) \ll (\bar{\sigma},N)$, the Hamiltonian can be written in terms of its matrix elements as
\begin{eqnarray}
H_{coll}^{SB} &\simeq& \sum_{n,y} \left|n,y\right> \left[ E_{coll}^0 + h y + \frac{2 J}{N} n^2 \right] \left<n,y\right| \nonumber \\
&&\hspace{8pt} - \left|n \pm 1,y\right> \left[ \frac{B}{4} f(y)  \right] \left<n,y\right|,
\end{eqnarray}
where $f(y) \equiv \left(2-\frac{y}{\bar{\sigma}}\right) \sqrt{\left(\frac{N}{2}\right)^2-\bar{\sigma}^2}$. If we write the eigenfunctions of this equation as $\left|x,y\right> = \sum_n \Psi \left( n,x \right) \left|n,y\right>$ and take the continuum limit, then Schr\"odinger's equation reduces to the well known harmonic oscillator equation,
\begin{eqnarray}
- \frac{1}{2} \frac{\partial^2}{\partial n^2} \Psi \left( n,x \right) + \frac{1}{2} \omega^2 n^2 \Psi \left( n,x \right) = \nu \Psi \left( n,x \right),
\end{eqnarray}
with $\omega^2 = \frac{8 J}{B N f(y)}$ and $\nu = 1+2\frac{E(x,y)-E_{coll}^0-h y}{B f(y)}$. The wavefunctions $\Psi$ are the eigenfunctions of the harmonic oscillator, which can be written explicitly in terms of Hermite polynomials. The corresponding eigenvalues obey $\nu = (x+1/2)\omega$, and thus we find the energies of the symmetry broken collective Hamiltonian~Eq.\eqref{HcollSB} to be given by
\begin{eqnarray}
E\left(x,y\right) &=& E_{coll}^0 + h y - \frac{1}{2} B N g(y) \nonumber \\
&&+ \left(x+\frac{1}{2}\right) \sqrt{2 J B} \sqrt{g(y)},
\end{eqnarray}
where $g(y) \equiv \left( 1 - 2\frac{y}{\bar{\sigma}} \right) \sqrt{1-\left(\frac{h}{2 J}\right)^2}$. The term $\propto BN$ in this expression shows that the symmetry of the system will be spontaneously broken: even if only an infinitesimally small symmetry breaking field is present, the pseudospins can gain an infinite amount of energy in the limit of $N \to \infty$ by aligning with that field. In the thermodynamic limit the alignment will thus happen spontaneously and the resulting symmetry broken state is exactly the expected canted antiferromagnet.

\section{Gauge Volume}
Having found the the eigenfunctions of the collective symmetry broken Hamiltonian, the question arises what these states represent, and even if they are truly physical states. As mentioned before, the symmetry breaking field in the collective Hamiltonian~Eq.\eqref{HcollSB} acts as an implicit gauge fix.
It is not a priori clear whether or not this (non-physical) gauge fixing introduced any extra unphysical states in the spectrum. If we define the gauge volume of a certain state to be the collection of all states that are connected to it by a gauge transformation, then making a specific gauge choice in the Hamiltonian can in principle lead to the erroneous identification of states within the same gauge volume as seperate physical states. The question is thus whether the excited states  of $H_{coll}^{SB}$ that we found are part of its ground state gauge volume or not.

\begin{figure}
\includegraphics[width=0.85\columnwidth]{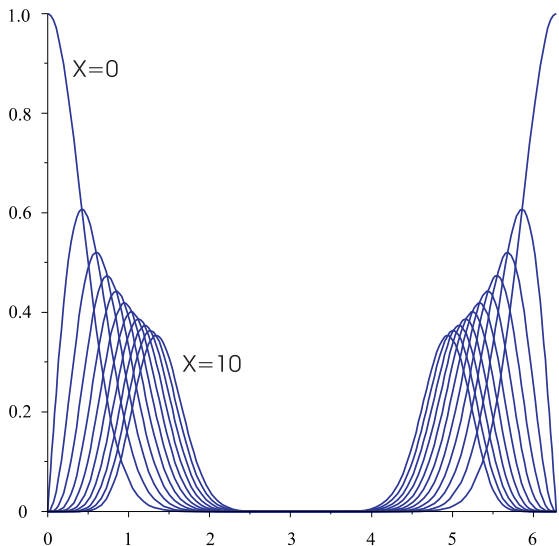}
\caption{The overlap between the thin spectrum state $\left| x \right>$ and
  the rotated groundstate $\hat{R}\left( \theta\right) \left| 0 \right>$, as a
  function of the angle of rotation $\theta$, for different values of $x$. To make this graph we used the values
  $J=10$, $B=h=1$ and $N=100$.}
\label{overlap}
\end{figure}

The ground state of the collective Hamiltonian
is an ordered antiferromagnet in terms of pseudospins, and we have seen that
it corresponds to a superconducting state of Cooper pairs. The excitations
labeled by $x$ in the pseudospin picture must involve the superposition of
collective excitations with wavenumbers $k=0$ and $k=\pi$. However as mentioned
before, the gauge volume of this system is made up of global uniform rotations
of the entire pseudospin lattice around the $z$-axis. Proving that these excitations are not
within the gauge volume of the groundstate wavefunction, therefore amounts to showing that the excited states cannot be written as only a global rotation of the groundstate. Using the explicit
formulas for the eigenfunctions of $H_{coll}^{SB}$ it is easy to check that
indeed the overlap between the state $\left| x = X \right>$ and the state
$\left| x=0 \right>$, rotated over an angle $\theta$, is one if and only if
both $X$ and $\theta$ are zero (see figure~\ref{overlap}). This proves that indeed the excited state cannot be
written only as a global rotation of the groundstate, and thus that the excited state is
not within the ground state's gauge volume.\footnote{The excitations labeled by $y$ were already identified as being additions (or subtractions) to the average total number of Cooper pairs in the superconductor. They alter the $z$ projection of the pseudospins and thus trivially shown to be outside of the groundstate's gauge volume.}

\section{Josephson Effect}
We have seen that an infinitesimally weak coupling between the local pairing superconductor and a second external superconductor gives rise to spontaneous symmetry breaking of the total phase difference between the two superconductors. In this section we show that this symmetry breaking is an essential prerequisite for the observation of the Josephson effect.  This effect is well-known to rely on the existence of a phase difference between two superconductors, in which case a finite (but weak) coupling between the two systems causes a supercurrent between them.\cite{Josephson62, Josephson69, Anderson64}

To study the Josephson effect in the local pairing model for superconductivity, we will follow the approach of Anderson by treating the hopping term in equation~Eq.\eqref{eq:Hsb} as a small perturbation and computing the change in energy due to it.\cite{Anderson64} By taking the derivative of the the first order correction to the energy with respect to the vector potential we can then directly obtain the supercurrent.

First we construct the state in which there are no excitations $y$ in either superconductor, and in which the thin spectrum state of both superconductors is $\left| x=0 \right>$, which we denote as $|0\rangle \equiv \left| 0,0 \right>_{\Sigma} \left| 0,0 \right>_{\sigma}$. Now we rotate the phase of the $\Sigma$-superconductor by an angle $\theta$ and get the state $|\theta\rangle= \hat{R}\left( \theta \right) |0\rangle$, where $\hat{R}$  rotates the state $\left| 0,0 \right>_{\Sigma}$ over an angle $\theta$ around the $z$-axis.  The fist order correction to the energy is
\begin{eqnarray}
\Delta E_1 &=&  T \left<  \theta \right|
(\Sigma_A^+ \sigma_A^- + \Sigma_B^+ \sigma_B^-)   \left| \theta   \right> + \text{H.c.} \nonumber \\
&=& T \frac{N_{\sigma} N_{\Sigma}}{4}  \cos \left( \theta \right) \nonumber \\
&& \sum_{n,n'} \Psi \left( n'+1,0 \right) \Psi \left( n-1,0 \right) \Psi \left( n',0 \right) \Psi \left( n,0 \right)\nonumber \\
&=&  \Delta E  \cos \left( \theta \right) ,
\label{E1}
\end{eqnarray}
The summation over the Hermite polynomial wavefunctions $\Psi\left(n,x\right)$ can easily be evaluated numerically and is of order unity. In the last line the energy change $\Delta E$ is implicitly defined. The expression above for the energy difference in terms of the phases of the transformed pseudospins $\sigma$ and $\Sigma$ is gauge invariant, because a gauge transformation rotates all pseudospins simultaneously, and leaves the phase difference $\theta$ invariant. As we wish to determine the derivative of $\Delta E_1$ with respect to the vector potential, we need to go back to the formulation of the problem in terms of the original $\bf S$ pseudospins. By inspection of~Eq.\eqref{E1} it is clear that the gauge-dependent expression for the energy change will be
\begin{eqnarray}
\Delta E_1 &=& \Delta E \cos \left( \tilde{\theta} - \frac{2e}{\hbar c}\int {\bf A} \cdot d{\bf l} \right),
\label{E1A}
\end{eqnarray}
where $\tilde{\theta}$ describes the phase difference in terms of the original pseudospins, and the integration runs over a line connecting the two superconductors. We can now directly find the supercurrent by taking the derivative of the total energy with respect to the vector potential. Doing so and then transforming back to the pseudospins $\sigma$ and $\Sigma$ yields
\begin{eqnarray}
\left< {J} \right> &=& c \frac{\delta \left< H \right>}{\delta {A}}= \frac{2 e \Delta E}{\hbar} \sin \left( \theta \right),
\label{J1}
\end{eqnarray}
which is precisely the expected Josephson current.\cite{Anderson64} Notice that the occurrence of the Josephson effect is a direct consequence of the phase symmetry breaking. Starting from a symmetric, total-number ground state for the superconductors, the first order correction to the energy induced by the hopping term vanishes, so that there is no Josephson current in that case.

This observation raises the question what happens in practice when we try to bring two isolated pieces of superconductor closer together and allow them to weakly couple. Assuming that the superconductors were so well isolated that their initial states were total number states, the first influence of the coupling will be to break the phase symmetry and cause the superconductors to acquire a well-defined phase {\it difference}. If two superconductors approach each other from infinity the actual value of the phase difference $\theta$ will be zero --due to the energy gain~Eq.\eqref{E1}-- and no Josephson current of the type~Eq.\eqref{J1} is present. However, a finite phase difference between the two pieces of superconductor can be induced by first applying a voltage difference between the two systems and then letting them approach. In the Hamiltonian the voltage bias can be introduced into the hopping term by the Peierls construction:
\begin{eqnarray}
T \left(e^{i \theta} \Sigma_A^+ \sigma_A^- + e^{i \theta} \Sigma_B^+ \sigma_B^- + \text{H.c.} \right),
\end{eqnarray}
where $\theta$ depends on the applied voltage. This form of the coupling term causes the symmetry to be broken such the phase difference between the superconductors is $\theta$. If the bias is switch off again, a  Josephson current as in~Eq.\eqref{J1} is induced.

\section{Decoherence}
We would now like to apply the results of the previous sections to the description of quantum coherence. In analogy to the result for antiferromagnets,\cite{vanWezel05} we expect the existence of the unobservable thin spectrum to give rise a maximum coherence time $t_{spon} \propto N \hbar / k_B T$.

Let us define a qubit made of the eigenstates of the collective part of the local pairing superconductor. If temperature is sufficiently low (i.e. $k_B T \ll J, h$) then we can use the states $y=0$ and $y=1$ as the computational states of such a qubit. These states correspond to states with a different number of Cooper pairs, and qubits of this type have been made experimentally in the form of Cooper-pair boxes.\cite{Nakamura99, Makhlin01,Nakamura02} In these Cooper-pair boxes a superconducting island can be brought into a superposition of having $\bar{N}$ and $\bar{N}+1$ Cooper-pairs present. Superpositions of this type can reach coherence times of up to 500 ns.\cite{Vion02,Siddiqi06}

In our local pairing description of the qubit, the initial state of the system
must be a thermal mixture of thin spectrum states. After all, controlling
these states experimentally is practically impossible. The initial state
should then be brought into some superposition of the computational states
$y=0$ and $y=1$, so that it can be used in a quantum computation. Because we
know all eigenstates and eigenvalues of the Hamiltonian exactly, we can then
explicitly follow the time evolution of the superposition.\cite{vanWezel06} The complete process is thus described by
the time dependent density matrix
\begin{eqnarray}
\rho_{t<0} &=& \frac{1}{Z} \sum_x e^{-\beta E\left(x,0\right)} \left| x,0 \right>\left< x,0\right| \nonumber \\
\rho_{t=0} &=& \frac{1}{2Z} \sum_x e^{-\beta E\left(x,0\right)} \left[ \left| x,0 \right> + \left| x,1 \right> \right] \left[ \left< x,0\right| +\left< x,1\right| \right]  \nonumber \\
\rho_{t>0} &=& \frac{1}{2Z} \sum_x e^{-\beta E\left(x,0\right)} \left[ \left| x,0 \right>\left< x,0\right| +\left| x,1 \right> \left< x,1\right| \right. \nonumber \\
&&\left. + e^{-\frac{i}{\hbar} \left(E\left(x,0\right) - E\left(x,1\right) \right) t} \left| x,0 \right> \left< x,1\right| + \text{h.c.}\right].
\label{matrices}
\end{eqnarray}
where $Z$ is the partition function at $t<0$. The thin spectrum states labeled with $x$ cannot be observed or controlled experimentally, and they should therefore be traced out of the final density matrix. The remaining reduced density matrix then shows the coherence of only the superposition of $y$ states. The disappearance of the off-diagonal matrix element of the reduced density matrix serves as a measure of the resulting coherence time, and it can easily be checked that this coherence time is given by
\begin{eqnarray}
t_{spon} = \frac{2 \pi \hbar}{k_B T} \frac{ \bar{\sigma}}{2}.
\label{tspon}
\end{eqnarray}
Here $\bar{\sigma}$ signifies, as before, the average number of Cooper pairs on the superconducting island in the groundstate. This coherence time is the maximum coherence time of a superconducting island, which is limited by the existence of a thin spectrum in the superconductor. Just as in the cases of crystals and antiferromagnets, the details of the model (e.g. $J$ or $h$) do not enter into the expression for the maximum coherence time, which thus looks like a universal timescale.\cite{vanWezel05,vanWezel06}

Filling in the values for the constants $\hbar$ and $k_B$ and taking $\bar{\sigma} \simeq 10^6$ and $T \simeq 40$ mK,\cite{Nakamura02} we find a maximum coherence time for the experimentally realized Cooper pair boxes of $\simeq 0.5$ ms.
Clearly this timescale set by the presence of the thin spectrum states which are associated with the spontaneous symmetry breaking, is much larger than the timescale that is the current limit to coherence of the Cooper-pair boxes due to other environmental factors. However, it is well possible that the limit set by the thin states will come within the experimental reach in the near future, either because the isolation from external sources of decoherence will be developed further, or because the size of the Cooper-pair box itself is reduced even more.

\section{BCS Superconductor}
In the previous sections we have shown that the superconductive groundstate is a state with a spontaneously broken U(1) symmetry. As a consequence the superconductor must have a thin spectrum of states that describe the collective excitations on top of the ground state. In the case of a local pairing model for superconductivity we have found an explicit expression for these thin states and we have shown how they can cause decoherence if we try to use a superconductive island as a qubit.

It could be argued that the local pairing model is somewhat pathological, and not really representative for real-life superconductors, even though from the point of view of symmetry the model is equivalent to a weak coupling model (because there is no phase transition which separates the two). We will therefore also work out the symmetry breaking and decoherence in a BCS description, and show that although the details of the picture change, the underlying physics is exactly equivalent, and in fact gives rise to the exact same conclusions regarding the thin spectrum and the timescale on which decoherence will set in. The draw-back of doing the calculation in a mean field BCS description is that it cannot be done in a manifestly gauge invariant way, so that the role of the vector potential is obscured.

After creating Cooper pairs, we arrive in the standard BCS theory at the effective Hamiltonian
\begin{eqnarray}
H_{BCS} &=& \sum_k \epsilon_k \left( c^{\dagger}_k c^{\phantom \dagger}_k + c^{\dagger}_{-k} c^{\phantom \dagger}_{-k} \right) \nonumber \\
&&- \sum_{k\neq k'} U ~ c^{\dagger}_k c^{\dagger}_{-k} c^{\phantom \dagger}_{-k'} c^{\phantom \dagger}_{k'}.
\label{Hbcs}
\end{eqnarray}
Here we have adopted the convention to write $(k,\uparrow)$ as $k$ and $(-k,\downarrow)$ as $-k$. The dispersion of the bare Fermi-sea is characterized by $\epsilon_k$ while $U$ is the effective pairing interaction due to phonon exchange. $U$ is non-zero and attractive only in a shell around the Fermi energy with a width of about the Debije energy. It is easy to see that extensivity of the model in fact requires $U$ to be inversely proportional to the total number of electrons in the system. We will therefore redefine the pairing potential as $U=V/N$, where $N$ denotes the total number of electrons in the $k$-space shell in which $U$ is non-zero.

By writing down the Hamiltonian~Eq.\eqref{Hbcs} we have assumed that there is no external magnetic field and we have fixed the gauge to ensure that the electromagnetic vector potential vanishes everywhere.
Anderson showed that the BCS Hamiltonian in this form can be rewritten as a spin problem by introducing the pseudospins\cite{Anderson58}
\begin{eqnarray}
S_k^+ &=& c^{\phantom \dagger}_{-k} c^{\phantom \dagger}_{k}  \nonumber \\
S_k^z &=& \frac{1}{2} \left[1 - c^{\dagger}_k c^{\phantom \dagger}_k - c^{\dagger}_{-k} c^{\phantom \dagger}_{-k} \right].
\end{eqnarray}
In the subspace without any quasiparticles (i.e. $n_k=n_{-k} ~\forall k$), the Hamiltonian up to an overall constant becomes
\begin{eqnarray}
H_{BCS} = - 2 \sum_k \epsilon_k S_k^z - \sum_{k\neq k'} \frac{V}{N} \left( S_k^x S_{k'}^x + S_k^y S_{k'}^y \right).
\label{Hspin}
\end{eqnarray}
\begin{figure}
\includegraphics[width=.7\columnwidth]{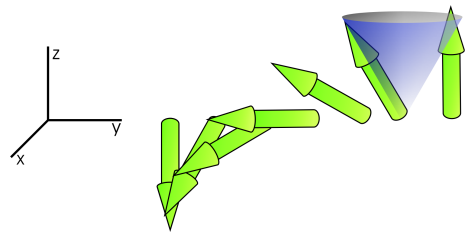}
\caption{A schematic representation of the region of width $k_D$ around
  $k_F$. The arrows represent the pseudospins ${\bf S}$. Spontaneous
  symmetry breaking causes the projections of the pseudospins in the $xy$-plane
  to align.}
\label{wall}
\end{figure}
Interpreted at face value, this Hamiltonian describes pseudo spins on a
lattice which has position-label $k$. On this lattice, three different and
independent regions can be identified. In the region $k < k_F - k_D$ (where
$k_F$ is the Fermi wavenumber and $k_D$ the Debije wavenumber) we know that
the pairing potential vanishes and $\epsilon_k$ is negative, so that all
pseudospins in that region will point down, which corresponds to completely
filled electronic states. In the region $k > k_F + k_D$ the pairing potential
is zero as well, but here $\epsilon_k$ will be positive, causing all spins to
point up, and all electronic states to be empty. In the shell of width $k_D$
around $k_F$ a more interesting situation occurs. There $V$ is nonzero (and
approximately constant), while $\epsilon_k$ switches sign right at $k_F$. The
pseudo-spin structure that one would classically expect in that region is that
of a magnetic domain wall: the pseudospins point up at one end of the region,
then continuously fall over until they reach the $xy$ plane exactly at $k_F$,
and then they continue on until they point down at the other
end (see figure~\ref{wall}). Electronically that structure corresponds to the BCS wavefunction
$\prod_k \left(u_k +v_k c^{\dagger}_k c^{\dagger}_{-k} \right) \left|
\text{vac} \right>$.

The Hamiltonian $H_{BCS}$ however is invariant under rotations around the
$z$-axis, and the exact groundstate will also obey this symmetry and have a
completely delocalized projection of the pseudospins on the $xy$ plane. To
form a true domain wall, and thus the classical superconducting state, this
U(1) symmetry will have to be spontaneously broken.

Because the symmetry breaking will only have an effect in the region around
$k_F$ and because this region is fully decoupled from the other two regions of
$k$-space, we will focus solely on that shell from now on, and define all sums
over $k$ to run from $k_F-k_D$ to $k_F+k_D$. The collective dynamics of the
system will again be described by the singular points of the Bogoliubov
transformation which diagonalizes the Hamiltonian. Because of the
ferromagnetic sign, the collective model in this case consists of only the
$k=0$ part of equation~Eq.\eqref{Hspin}:
\begin{eqnarray}
H_{coll} &=& - \frac{2}{N} \epsilon_{tot}^{\phantom z} S_{tot}^z - \frac{V(N-1)}{N^2} \left[ S_{tot}^x S_{tot}^x + S_{tot}^y S_{tot}^y \right] \nonumber \\
 &\simeq& -\frac{V}{N} \left[ {\bf S}_{tot}^{\phantom z} \cdot {\bf S}_{tot}^{\phantom z} - S_{tot}^z S_{tot}^z \right],
\label{HcollSC}
\end{eqnarray}
where ${\bf S}_{tot} \equiv \sum_k {\bf S}_k$ and where in the last line we have neglected terms of order $1/N^2$ and set $\epsilon_{tot} = 0$. The latter can be thought of as a strong coupling approximation, in the sense that the Hamiltonian~Eq.\eqref{HcollSC} will certainly be relevant in the region where $\epsilon_{tot} \ll V$. We will discuss different approximations for $\epsilon_{tot}$ at the end of this section. The eigenstates of the collective Hamiltonian are trivially found to be labeled by the total spin quantum number $S$ and its $z$-projection $M$, while the corresponding energies are given by $E_{coll}(S,M)=-V/N\left(S\left(S+1\right)-M^2\right)$.
The thin spectrum in this case is labeled by $M$, and describes states with different total electron densities. The total spin excitations labeled by $S$ on the other hand, are gapped with an energy $\sim V$. To break the $xy$-symmetry of $H_{coll}$ we can add a symmetry breaking field $-B S_{tot}^x$ along for example the $x$-axis. After evaluating its matrix elements\cite{vanWezel06} and taking the continuum limit, Schr\"odinger's equation can once again be written as a harmonic oscillator equation
\begin{eqnarray}
- \frac{1}{2} \frac{\partial^2}{\partial M^2} \Psi \left( M,x \right) + \frac{1}{2} \omega^2 M^2 \Psi \left( M,x \right) = \nu \Psi \left( M,x \right),
\end{eqnarray}
with $\omega^2=\frac{2V}{BNS}$ and $\nu=1+\frac{E(S,x)-E_{coll}(S,0)}{BS}$. The symmetry broken wavefunctions $\left|S,x\right> \equiv \sum_M \Psi(M,x) \left|S,M\right>$ thus have energies
\begin{eqnarray}
E(S,x) &=& -\frac{V}{N} S\left(S+1\right) - B S \nonumber \\
&&+ \left(x+\frac{1}{2}\right) \sqrt{V B} \sqrt{\frac{2S}{N}}.
\end{eqnarray}

In the ground state $S$ will be maximal (i.e. $N/2$), and then the term $\propto NB$ in the energy signals spontaneous symmetry breaking: in the thermodynamic limit the system can gain an infinite amount of energy by aligning with an infinitesimally small symmetry breaking field. The collective excitations that make up the (dual) thin spectrum on top of the symmetry broken ground state are labeled by $x$. Their energies are slightly influenced by the remaining collective quantum number $S$. If we make a superposition of total spin states and trace away the unobservable thin spectrum, then this small shift in the thin spectrum's energy levels will cause the decoherence of the visible reduced density matrix, in a manner completely analogous to the one described in equation~Eq.\eqref{matrices}. The resulting maximum coherence time is given by
\begin{eqnarray}
t_{spon} = \frac{2 \pi \hbar}{k_B T} N,
\label{tBCS}
\end{eqnarray}
where $N$ counts the number of states in the $k$ space volume of $k_D$ around $k_F$, which is proportional to the number of Cooper pairs in the superconducting condensate. So we find the same universal form for the expression of the coherence time set by spontaneous symmetry breaking as in the case of the local pairing model for superconductivity.

As mentioned before, the collective Hamiltonian~Eq.\eqref{HcollSC} can be seen as a strong coupling limit, because we require $\epsilon_{tot}$ to be much smaller than $V$. We can drive the system to a somewhat weaker coupling regime by reincluding an approximate form of $\sum_k \epsilon_k S_k^z$ into $H_{coll}$. One possible choice for such a term would be $t\left(S_{k_{min}}^z - S_{k_{max}}^z\right)$, which acts as a boundary condition, pulling the pseudospins down at the low $k$ boundary and up at the other end. A second choice could be the inclusion of the term $t\left(S_{A}^z - S_{B}^z\right)$ where $S_A$ consists of all spins with $k<k_F$ and $S_B$ denotes spins above the Fermi surface. In the latter case we should take care that $t$ cannot be too great, for if it would dominate over $V$ everywhere, then it would transform the domain wall structure of the superconducting state into a trivial Fermi-sphere structure again. It turns out that after some elaborate algebra  both of the above cases give the exact same form for the thin spectrum and the maximum coherence time as the ``bare'' model $H_{coll}$ did.

\section{Conclusions}
In this paper we have shown that the non-commuting observables of number and phase in a superconductor give rise to spontaneous symmetry breaking and an associated thin spectrum. We have given explicit expressions for these thin spectrum states in an array of Josephson junctions, in a tight-binding, negative-$U$ Hubbard model, and in the BCS model for superconductivity. Using the negative-$U$ Hubbard model we have commented on the relation between spontaneous symmetry breaking and its associated thin spectrum, and the Meissner effect, the Anderson-Higgs mechanism and the occurrence of Josephson currents. For the occurrence of the Meissner effect the phase symmetry actually need not be broken, but for the Josephson effect it does. We have also given a description of a gedanken experiment in which the superconductor is to be used as a qubit, and we have shown that the presence of the thin spectrum states associated with spontaneous symmetry breaking will lead to decoherence of the qubit within the time $t_{spon}=2 \pi \hbar N / k_B T$, where $N$ counts the number of Cooper pairs involved. This result was obtained in the negative-$U$ Hubbard model as well as the BCS model. The timescale $t_{spon}$ is universal in the sense that it does not depend on the underlying model parameters. Its form coincides precisely with that of the decoherence time induced by thin spectrum dynamics in antiferromagnets and quantum crystals.

The maximum coherence time that we found here for superconducting devices should apply directly to experimental realizations of the so called Cooper pair box, and thus give a maximum coherence time of the order of milliseconds. The decoherence caused by the thin spectrum at the moment is much weaker than that caused by other sources, but it may well come within experimental reach in the near future. To apply the results of this paper to other types of superconducting qubits, such as for example superconducting flux qubits, one should adjust the models used here in order to also accommodate for the presence of an external magnetic flux and an associated supercurrent in the groundstate. Because of its universal nature it is expected that the  decoherence time set by the thin spectrum in these cases also will be given by the timescale $t_{spon}=2 \pi \hbar N / k_B T$.

\section{Acknowledgements}
We thank Jan Zaanen
for numerous stimulating discussions and gratefully acknowledge support from the Dutch Science Foundation FOM.


\end{document}